\documentclass[aps,prl,twocolumn,showpacs,amsmath,amssymb,floatfix,superscriptaddress]{revtex4-1}
\usepackage{graphicx}

\begin{document}

\title{A simple screened exact-exchange approach for excitonic properties in solids}
\author{Zeng-hui Yang}
\affiliation{Department of Physics, Temple University, Philadelphia, PA 19122, USA}
\affiliation{Department of Physics and Astronomy, University of Missouri, Columbia, MO 65211, USA}
\author{Francesco Sottile}
\affiliation{Laboratoire des Solides Irradi\'{e}s, \'{E}cole Polytechnique, CNRS, CEA-DSM, F-91128 Palaiseau, France}
\affiliation{European Theoretical Spectroscopy Facility (ETSF)}
\author{Carsten A. Ullrich}
\affiliation{Department of Physics and Astronomy, University of Missouri, Columbia, MO 65211, USA}
\date{\today}
\pacs{31.15.ee, 
71.15.Qe,	
71.35.Cc, 	
78.20.Bh	
}

\newcommand{\Hxc}{_{\rm Hxc}}
\newcommand{\Har}{_{\rm H}}
\newcommand{\xc}{_{\rm xc}}
\newcommand{\br}{{\bf r}}
\newcommand{\bfk}{{\bf k}}
\newcommand{\bfG}{{\bf G}}
\newcommand{\vect}[1]{\mathbf{#1}}
\newcommand{\parref}[1]{(\ref{#1})}
\newcommand{\matelem}[3]{\left\langle #1 \left| #2 \right| #3 \right\rangle}

\begin{abstract}
We present a screened exact-exchange (SXX) method for the efficient and accurate calculation of the optical properties of solids,
where the screening is achieved through the zero-wavevector limit of the inverse dielectric function.
The SXX approach can be viewed as a simplification of the Bethe-Salpeter equation (BSE) or, in the context of time-dependent
density-functional theory, as a first step towards a new class of hybrid functionals for the optical properties of solids.
SXX performs well for bound excitons and continuum spectra in both small-gap semiconductors and large-gap insulators,
with a computational cost much lower than that of the BSE.
\end{abstract}

\maketitle

The Bethe-Salpeter equation (BSE) \cite{HS80,ORR02} is considered the gold standard for calculating the optical properties of periodic solids and
many other materials. The nonempirical nature of the BSE guarantees its wide applicability and high degree of accuracy, but its computational cost
becomes prohibitive beyond the simplest systems. Time-dependent density-functional theory (TDDFT) \cite{RG84,MMNG12,U12} is computationally much less expensive, and is therefore a popular alternative for the calculation of optical properties.  TDDFT calculations can be orders of magnitude faster than the BSE, but none of the existing empirical or nonempirical exchange-correlation (xc) kernels for solids \cite{Sottile2003,BSVO04,SDSG11,TTCO13,deBoeij2001} can achieve the same level of accuracy for both small-gap and wide-gap solids.
The exception is the so-called ``nanoquanta'' xc kernel \cite{ORR02,Reining2002,SOR03,ADM03,MDR03}, which is as accurate as the BSE, but equally expensive.

Recent TDDFT studies for solids have identified the crucial importance of the long-range part of the xc kernel \cite{GGG97,Botti2007}.
Exact-exchange TDDFT \cite{Kim2002a,Kim2002b} successfully produces excitonic properties, but the Coulomb singularity needs
to be cut off, which is equivalent to using a screened Coulomb interaction \cite{Bruneval2006}.
Hybrid xc functionals are defined as a mixture of semilocal (gradient-corrected) xc functionals with a fraction
of nonlocal exact exchange. The B3LYP hybrid functional \cite{SDCF94} has been used
to calculate optical spectra in semiconductors \cite{Bernasconi2011,Tomic2014}, with a generally good description of
optical gaps, despite the fact that the 0.2 mixing parameter of B3LYP is optimized for finite systems.
The HSE functional \cite{Heyd2003,Heyd2006} uses exact exchange for short-range interactions only; this
produces very good quasiparticle gaps \cite{Heyd2005,Schimka2011,Moussa2012}, but HSE cannot yield bound excitons, although it may still give
decent continuum spectra  \cite{Paier2008}.

In this paper we propose a simple, nonempirical and material-dependent way of screening the long-range Coulomb exchange,
which can be viewed as a simplified BSE approach. We show that this screened exact-exchange (SXX) approach outperforms all
TDDFT approaches currently on the market, retaining most of the accuracy of the BSE over a wide range of materials, but at a much lower computational cost. This builds a bridge between TDDFT and many-body theories, and opens up new directions towards the development of hybrid functionals for the optical properties of insulating solids.

Although TDDFT and BSE are very different theories, the excitation spectra in solids are in both cases obtained through an eigenvalue equation \cite{C96,ORR02} (atomic units [$e=\hbar=m_e=1/4\pi\epsilon_0=1$] are used throughout):
\begin{eqnarray}
\sum_{(m\bfk_mn\bfk_n)}\Big[\delta_{i\bfk_i,m\bfk_m}\delta_{j\bfk_j,n\bfk_n}
(\epsilon_{j\bfk_j}-\epsilon_{i\bfk_i}) && \nonumber\\
\left. {} +F\Hxc^{(i\bfk_ij\bfk_j)(m\bfk_mn\bfk_n)}\right]\rho^{(m\bfk_mn\bfk_n)}_\lambda
&=&
\omega_\lambda\rho^{(i\bfk_ij\bfk_j)}_\lambda,
\label{eqn:Casida}
\end{eqnarray}
where $i$ and $m$ denote occupied bands, $j$ and $n$ denote unoccupied bands, the $\epsilon$'s are single-particle energies (either quasiparticle
or Kohn-Sham), and $\omega$ is the excitation frequency. The main difference lies in the coupling matrix $F\Hxc=F\Har+F\xc$. For optical properties, only vertical excitations are considered, so that $\bfk_i=\bfk_j$ and $\bfk_m=\bfk_n$ in Eq. \parref{eqn:Casida}. The Hartree part of the coupling matrix is the same in the two methods, and is given by
\begin{eqnarray}
F\Har^{(ij\bfk)(mn\bfk')}&=& \frac{2}{V_\text{crys}}\sum_{\bfG\ne 0}\frac{4\pi}{|\bfG|^2}
\matelem{j\vect{k}}{e^{i\vect{G}\cdot\vect{r}}}{i\vect{k}}
\nonumber\\
&\times&
\matelem{m\vect{k'}}{e^{-i\vect{G}\cdot\vect{r}}}{n\vect{k'}}.
\label{eqn:FH}
\end{eqnarray}
The long-range part ($\vect{G}=0$) of the Coulomb interaction is omitted so that the eigenvalues of Eq. \parref{eqn:Casida} correspond to poles in the macroscopic dielectric function \cite{ORR02,UY14}.

For the BSE, as well as for our SXX method, the xc part of the coupling matrix can be written as
\begin{eqnarray}
\lefteqn{
F\xc^{(ij\vect{k})(mn\vect{k}')}=\frac{1}{V_\text{crys}}\sum_{\vect{G}\vect{G}'}g_{\vect{G}\vect{G}'}(\vect{q})} \nonumber\\
&&
\times\matelem{j\vect{k}}{e^{i(\vect{q}+\vect{G})\cdot\vect{r}}}{n\vect{k}'}\!
\matelem{m\vect{k}'}{e^{-i(\vect{q}+\vect{G})\cdot\vect{r}}}{i\vect{k}}\delta_{\vect{q},\vect{k}-\vect{k}'}.
\label{eqn:FxcTDHFBSE}
\end{eqnarray}
Here, $g_{\vect{G}\vect{G}'}(\vect{q})=-4\pi\gamma\delta_{\vect{G}\vect{G}'}/|\vect{q}+\vect{G}|^2$ for SXX ($\gamma$
is a screening parameter, to be further specified below), and $g_{\vect{G}\vect{G}'}(\vect{q})=-4\pi\epsilon^{-1}_{\vect{G}\vect{G}'}(\vect{q},\omega=0)/|\vect{q}+\vect{G}'|^2$ for the BSE.
$\epsilon^{-1}$ is the inverse dielectric function, obtained within the random phase approximation (RPA) as
\begin{equation}
\epsilon^{-1}_{\vect{G}\vect{G}'}(\vect{q},\omega)=\delta_{\vect{G}\vect{G}'}+\frac{4\pi}{|\vect{q}+\vect{G}|^2}\chi^\text{RPA}_{\vect{G}\vect{G}'}(\vect{q},\omega),
\end{equation}
with the RPA response function defined as $\chi^\text{RPA}=\chi_0+\chi_0 v \chi^\text{RPA}$ ($\chi_0$ is the quasiparticle response function \cite{U12}).

In TDDFT, the xc part of the coupling matrix is
\begin{eqnarray}
F\xc^{(ij\vect{k})(mn\vect{k}')} &=&
\frac{2}{V_\text{crys}}\sum_{\vect{G}\vect{G}'}f_{{\rm xc},\vect{G}\vect{G}'}(\vect{q}=0) \nonumber\\
&\times&
\matelem{j\vect{k}}{e^{i\vect{G}\cdot\vect{r}}}{i\vect{k}}\matelem{m\vect{k'}}{e^{-i\vect{G}'\cdot\vect{r}}}{n\vect{k'}},
\label{eqn:Fxc}
\end{eqnarray}
where $f\xc(\br,\br')=\delta v\xc(\br)/\delta n(\br')$ is the adiabatic xc kernel of TDDFT. The structure of Eq. \parref{eqn:Fxc} is similar to Eq. \parref{eqn:FH}, but different from Eq. \parref{eqn:FxcTDHFBSE}: only the $\vect{q}=0$ part of $f\xc$ enters in the expression, so the head ($\vect{G}=\vect{G}'=0$) at $\vect{q}=0$ of the xc kernel plays a much more important role in TDDFT than $g_{00}(0)$ in the BSE.

To illustrate the difficulty in developing a universally applicable nonempirical xc kernel,
we consider the long-range corrected (LRC) kernel \cite{GGG97,BSVO04,BSOR06}
\begin{equation} \label{LRC}
f^\text{LRC}_{\rm xc,\vect{G}\vect{G}'}(\vect{q})=-\frac{\alpha}{|\vect{q}+\vect{G}|^2} \, \delta_{\vect{G}\vect{G}'},
\end{equation}
which represents the long-range part of
the exact xc kernel in insulators. The empirical parameter $\alpha$ acts as a rough approximation to the dielectric screening effects within the BSE, but has no clear justification in the TDDFT framework, besides giving the correct asymptotic behavior. In Ref. \cite{BSVO04},
the relation $\alpha = 4.615 \, \epsilon_\infty^{-1}-0.213$ was proposed, which works quite well for the optical spectra of semiconductors, but fails for insulators.

In Ref. \cite{YU13}, $\alpha$ was fitted against experimental exciton binding energies for various materials. It was found that the value of $\alpha$ spans a wide numerical range, from 0.595 (GaAs) to 96.5 (solid Ne). For small-gap materials, the relative change in the exciton binding energy caused by a change in $\alpha$ is substantial; for large-gap materials, the exciton binding energies are not as sensitive. This shows how difficult it is to develop a widely applicable nonempirical xc kernel within the TDDFT framework \cite{YLU12}.

The situation is different in SXX. Unscreened time-dependent Hartree-Fock (TDHF) always overbinds excitons, so $\gamma$ has to be in the $[0,1]$ range for the correction to be in the right direction. Therefore, it is a much easier task to develop nonempirical approximations for $\gamma$  than for the TDDFT parameter $\alpha$.

To derive the SXX screening parameter $\gamma$, we start from the self-energy $\Sigma$:
\begin{eqnarray}
\lefteqn{
\Sigma(\br,\br',\omega)=\frac{i}{2\pi}G(\br,\br',\omega)W(\br,\br',0)} \nonumber\\
&=& \frac{i}{2\pi}G(\br,\br',\omega)\int d^3r''\;\epsilon^{-1}(\br,\br'',0)v(\br''-\br'), \label{eqn:Sigma}
\end{eqnarray}
where $G$ is the quasiparticle Green's function. $\epsilon^{-1}$ in Eq. \parref{eqn:Sigma} is the full dielectric screening of the BSE, and we want to find a way to average its effect and motivate replacing it with a constant. Assuming that we can replace the static nonlocal screening $\epsilon^{-1}(\br,\br')$ with a frequency-dependent uniform screening $\epsilon_\text{uni}^{-1}(\br-\br',\omega)$, we write
\begin{equation}
\Sigma(\br,\br',\omega)=\frac{i}{2\pi}\tilde{G}(\br,\br',\omega)\int d^3r''\;\epsilon_\text{uni}^{-1}(\br-\br'',\omega)v(\br''-\br'),
\label{eqn:Sigmauni}
\end{equation}
which defines the function $\tilde G$. Combining  Eqs. \parref{eqn:Sigma} and \parref{eqn:Sigmauni} in reciprocal space leads to
\begin{eqnarray}
\lefteqn{
\sum_{\vect{G}_2}\epsilon^{-1}_{\vect{G}_2 \vect{G}_1}(\vect{q}',0)G_{\vect{G}-\vect{G}_2,\vect{G}-\vect{G}_1}(\vect{q}-\vect{q}',\omega)}
\nonumber\\
&\equiv& \epsilon_{\text{uni},\vect{G}_1}^{-1}(\vect{q}',\omega)\tilde{G}_{\vect{G}-\vect{G}_1,\vect{G}'-\vect{G}_1}(\vect{q}-\vect{q}',\omega).
\label{eqn:epsilonG}
\end{eqnarray}
Eq. \parref{eqn:epsilonG} holds for any $\vect{q},\vect{G},\vect{G}'$. Setting these to zero and assuming the functions $G$ and $\tilde{G}$ to be real, we have
\begin{equation}
\epsilon_{\text{uni},\vect{G}}^{-1}(\vect{q},\omega)=\frac{\sum_{\vect{G}'}G_{\vect{G}\vect{G}'}(\vect{q},\omega)\epsilon^{-1}_{\vect{G}'\vect{G}}(\vect{q},0)}{\tilde{G}_{\vect{G}\vect{G}}(\vect{q},\omega)}.
\label{eqn:epsuni}
\end{equation}
In the $q\to0$ limit, Eq. \parref{eqn:epsuni} becomes
\begin{equation}
\lim_{q\to0}\epsilon_{\text{uni},\vect{G}}^{-1}(\vect{q},\omega)
=\frac{G_{\vect{G}0}(\vect{q},\omega)\epsilon^{-1}_{0\vect{G}}(\vect{q},0)}{\tilde{G}_{\vect{G}\vect{G}}(\vect{q},\omega)} \:.
\label{eqn:epsuniq0}
\end{equation}
For $\vect{G}\ne0$, Eq. \parref{eqn:epsuniq0} behaves like $O(1/q)$, so it is impossible to derive a general uniform screening as an average of the nonlocal screening of the BSE. However, the head ($\vect{G}=\vect{G}'=0$) of Eq. \parref{eqn:epsuniq0} has the correct $q\to0$ behavior, and can be used to approximate the BSE. Assuming that $\tilde{G}=G$, we obtain
\begin{equation}
\epsilon^{-1}_{\text{uni},0}(\vect{q},\omega)=\epsilon^{-1}_{00}(\vect{q},0).
\end{equation}
Now we make a rather drastic approximation by setting $\vect{q}=0$, since the long-range interaction dominates in solids. The screening parameter $\gamma$ then becomes
\begin{equation}
\gamma=\epsilon^{-1}_{00}(0,0),
\label{eqn:gamma}
\end{equation}
which is also the inverse of the infinite-frequency dielectric constant, $\epsilon_\infty^{-1}$, since phonon effects are not included. A similar simplified screening was proposed in Ref. \cite{MVOR11} for the nonlocal exchange part of a hybrid xc functional in order to obtain good band structures; by contrast, the purpose of our SXX is to yield good optical properties. We use the RPA for $\epsilon^{-1}_{00}(0,0)$ in actual calculations. Since $\epsilon=1-v\chi $, and the static $\chi$ at zero wave vector is negative, $\gamma$ of Eq. \parref{eqn:gamma} is bounded in the $[0,1]$ range.

Let us now compare the SXX approach with the BSE. The main difference is that in BSE the exchange is screened by the full inverse dielectric function $\epsilon^{-1}$, which makes it much more costly than SXX, where the screening parameter $\gamma$ is just a constant.
In practice, a BSE calculation is a four-step procedure: (i) ground-state calculation with a diagonalization over the selected $k$-points grid (often shifted \cite{Shirley1998} for optical properties); (ii) quasi-particle correction, typically within the GW
approximation \cite{Hedin1965}, but substituted with a simple scissor correction step in this work since we focus on the excitonic effects; (iii) generation of static screening $\epsilon^{-1}_{\vect{G}\vect{G}'}(\vect{q},\omega=0)$ within the RPA; (iv) construction and diagonalization of the excitonic Hamiltonian [i.e., Eq. (\ref{eqn:Casida})] containing the ingredients listed above.

\begin{figure}
\includegraphics[width=\columnwidth]{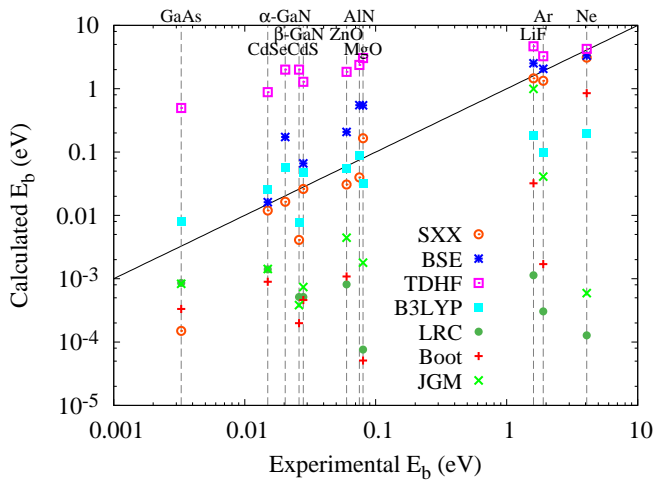}
\caption{(Color online) Comparison of calculated and experimental exciton binding energies $E_b$ in various materials (see Table \ref{table1} for further details). The solid line indicates where the calculated and experimental values of $E_b$ coincide.}
\label{fig1}
\end{figure}

Regarding  the computational workload, even though step (iv) has the worst scaling, step (iii) is often the most cumbersome part of the whole procedure, especially when one is interested in the small energy region of the spectrum and uses
the scissor operator procedure: the number of $q$-vectors in the screening is proportional to the number of $k$-points (since $\vect{q}=\vect{k}-\vect{k}'$) even for optical properties. This can become very demanding when many $k$-points have to be used together with many $\vect{G}$-vectors, as is the case for lower-dimensional systems, in particular 2D \cite{cudazzo}. In addition, the numerical evaluation of $\epsilon^{-1}_{\vect{G}\vect{G}'}(\vect{q},0)$ has a much worse convergence with the empty bands than the evaluation of the spectrum (for instance, the screening for LiF requires 20 empty bands, while the first exciton peak requires only one empty band). Our SXX approach bypasses this step and thus avoids a severe computational bottleneck in the description of optical properties at BSE-level for complex materials.

\begin{table*}
\begin{tabular}{cccccccccccc}
\hline\hline
 & GaAs & $\beta$-GaN & $\alpha$-GaN & CdS & CdSe & Ar & Ne & LiF & AlN & ZnO & MgO\\
\hline
Exp. & 3.27 & 26.0 & 20.4 & 28.0 & 15.0 & $1.90\times 10^3$ & $4.08 \times 10^3$ & $1.6 \times 10^3$ & 75 & 60 & 80\\
\\
BSE & --- & --- & 172  & 66.0  & 16.2  & $2.07\times 10^3$ & $3.32 \times 10^3$ & $2.51 \times 10^3$ & 552 & 208 & 546 \\
TDHF & 497 & $1.99 \times 10^3$ & $2.00\times 10^3$ & $1.28\times 10^3$ & 879 & $3.27\times 10^3$ & $4.26\times 10^3$ & $4.68\times 10^3$ & $2.37 \times 10^3$ & $1.84 \times 10^3$& $3.04 \times 10^3$\\
B3LYP\footnotemark[1] & 0.792  & 7.71  & 57.4  & 48.1 & 25.8 & 100 & 197 & 180 & 89.4 & 55.8 & 31.9\\
SXX & 0.151  & 4.08  & 16.4  & 26.0  & 11.9  & $1.33 \times 10^3$ & $3.08 \times 10^3$ & $1.46 \times 10^3$ & 39.9  & 30.8  & 165 \\
\\
LRC\footnotemark[2] & 0.858  & 0.514 & 0 & 0.513 & 1.40 & 0.304 & 0.127 & 1.14 & 0 & 0.810 & 0.076\\
Boot\footnotemark[3] & 0.332 & 0.199 & 0 & 0.461 & 0.895 & 1.70\footnotemark[4] & 852\footnotemark[4] & 32.2\footnotemark[4] & 0 & 1.09 & 0.051\\
JGM\footnotemark[5] & 0.833 & 0.382 & 0 & 0.741 & 1.42 & 41.0 & 0.593 & 993 & 0 & 4.45 & 1.79\\
\hline\hline
\end{tabular}
\caption{Exciton binding energies calculated with Eq. (\ref{eqn:Casida}), compared with experimental values (all numbers in meV). All calculations are head-only; see text for other technical details. The BSE results for GaAs and $\beta$-GaN were not calculated. The estimated error due to the head-only approximation is $<10$\% for all many-body calculations, and $<5$\% for TDDFT.}
\footnotetext[1]{Head-only calculation, equivalent to SXX with $\gamma=0.2$ independent of the material.}
\footnotetext[2]{With the empirical formula of Ref. \cite{BSVO04}.}
\footnotetext[3]{The bootstrap kernel of Ref. \cite{SDSG11}.}
\footnotetext[4]{The convergence of the bootstrap kernel strongly depends on the number of bands included in the iterative calculation of the kernel. These results are obtained by evaluating the bootstrap kernel with 30 bands. The results reported in Ref. \cite{YU13} were not fully converged.}
\footnotetext[5]{The jellium-with-gap model of Ref. \cite{TTCO13}.}
\label{table1}
\end{table*}

We have calculated the exciton binding energies $E_b$ of various semiconductors and insulators with SXX and other methods; the results are collected in Table \ref{table1} and in Fig. \ref{fig1}. Clearly, SXX produces a much better overall agreement with experiment than all TDDFT methods (except for GaAs), and yields an accuracy that is comparable to BSE across the board.

\begin{figure}
\includegraphics[width=\columnwidth]{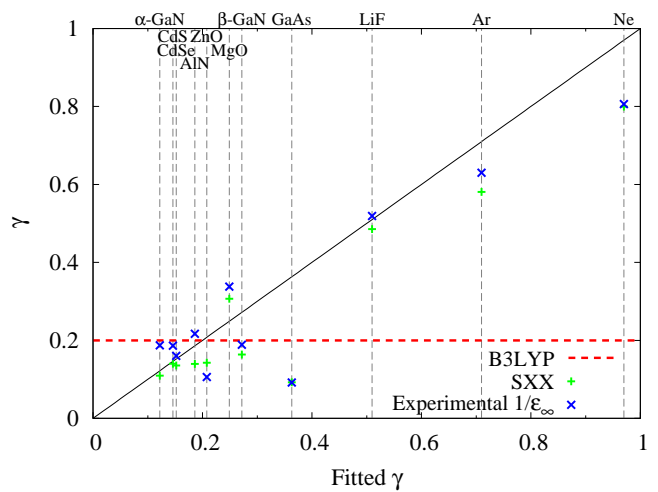}
\caption{(Color online) SXX screening parameter $\gamma$ [Eq. \parref{eqn:gamma}] and experimental value of $\epsilon_\infty^{-1}$ versus the fitted $\gamma$ reproducing the first exciton, for various materials. B3LYP corresponds to a constant value of $\gamma=0.2$. }
\label{fig2}
\end{figure}

The calculations of $E_b$ were done with the method described in Ref. \cite{YU13}. We use a scissor-corrected LDA ground state (calculated with ABINIT \cite{ABINIT}) as starting point. All calculations use three valence bands and one conduction band for Eq. \parref{eqn:Casida}, which is sufficient when only the exciton binding energies are of interest (but is, generally, insufficient
for the continuum part). We use an $18\times18\times18$ Monkhorst-Pack grid \cite{Monkhorst1976} for GaAs and $\beta$-GaN, a $15\times15\times15$ grid for MgO, a $10\times10\times10$ grid for Ar, Ne, and LiF, and a $20\times20\times20$ grid for other materials. To speed up the calculation, we only use the head of the xc kernel when calculating the coupling matrices [i.e., we neglect local-field effects by not taking the $\vect{G}$ and $\vect{G}'$ sums in Eq. \parref{eqn:FxcTDHFBSE} and \parref{eqn:Fxc}]. Including local-field effects generally changes $E_b$ very little (at most $\sim $10\%). Since the $E_b$ are already small numbers, the results without local-field effects are sufficiently accurate for our purposes. For the calculation of $\epsilon^{-1}_{00}(0,0)$, we use 60 bands for GaAs, $\beta$-GaN and MgO, and 30 bands for all other materials. 59 $\vect{G}$-vectors are used for $\epsilon^{-1}$, and the error for the convergence of $\epsilon^{-1}_{00}(0,0)$ is less than 1\%.

Figure \ref{fig2} compares $\gamma$ from Eq. \parref{eqn:gamma} with values of $\gamma$ fitted to reproduce the lowest experimental exciton binding energies. Aside from a few outliers (such as GaAs), the calculated and the fitted values of $\gamma$ are very close,
which explains the good performance for exciton binding energies in Table \ref{table1}. The experimental $\epsilon_\infty^{-1}$ values are also plotted in Fig. \ref{fig2}, showing that $\epsilon^{-1}_{00}(0,0)$ at the RPA level is already a good approximation to $\epsilon_\infty^{-1}$. It should be noticed that the B3LYP hybrid kernel \cite{SDCF94} (only the long-range part, which corresponds to $\gamma=0.2$, since the calculation only uses the head of the xc kernel) also performs well for semiconductors: Fig. \ref{fig2} shows that $\gamma=0.2$ is roughly the average of the semiconductor screening parameters. The B3LYP functional was designed with small molecules in mind, so its good performance for bound excitons in semiconductors seems fortuitous.

\begin{figure}
\includegraphics[width=\columnwidth]{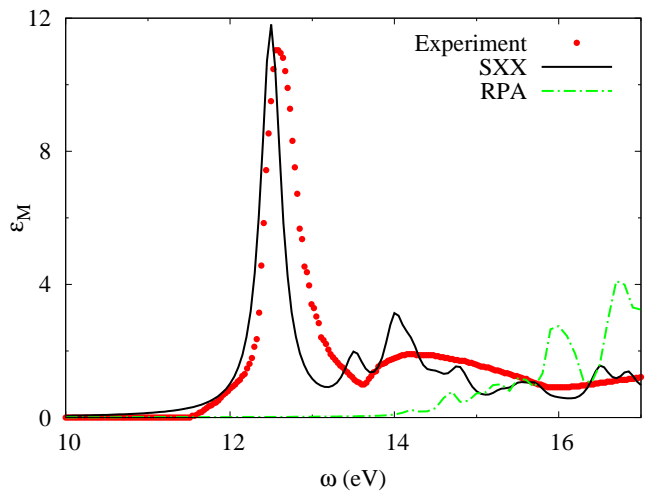}
\caption{(Color online) Absorption spectrum of LiF calculated with SXX and RPA, compared with experiment \cite{RW67}.}
\label{fig3}
\end{figure}

\begin{figure}
\includegraphics[height=\columnwidth,angle=-90]{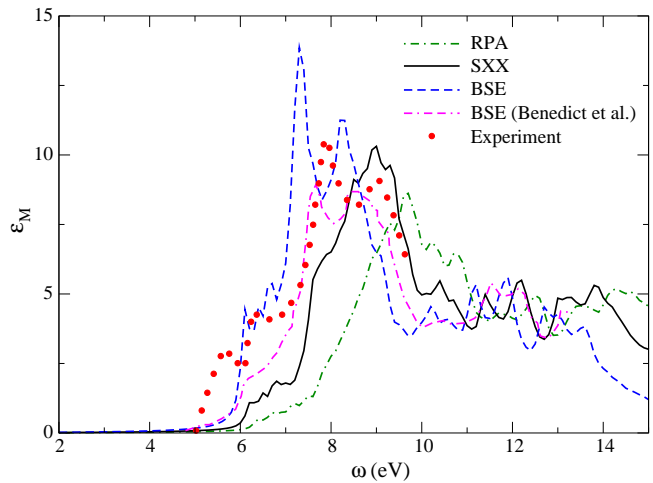}
\caption{(Color online) Absorption spectrum of AlN calculated with SXX, RPA, and BSE, compared with experiment \cite{CLKG05}. The BSE spectrum of Benedict {\em et al.} \cite{BWWC99} is also shown.}
\label{fig4}
\end{figure}

\begin{figure}
\includegraphics[height=\columnwidth,angle=-90]{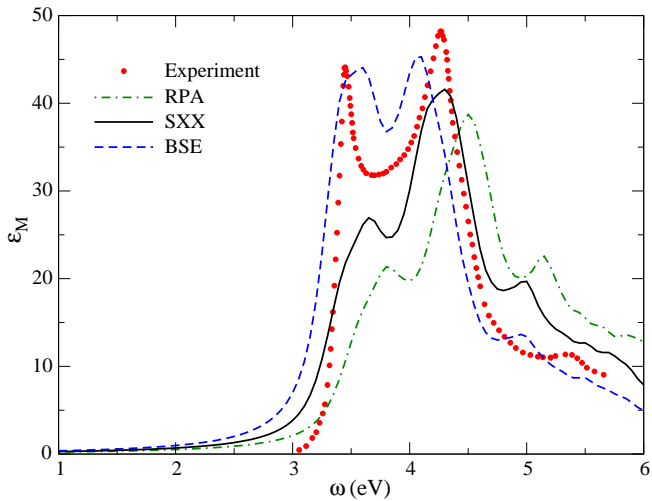}
\caption{(Color online) Absorption spectrum of Si calculated with SXX, RPA, and BSE, compared with experiment \cite{LGVC87}.}
\label{fig5}
\end{figure}

To demonstrate that our method yields good results not only for the exciton binding energies, we present the optical spectra of LiF (Fig. \ref{fig3}), AlN (Fig. \ref{fig4}), and Si (Fig. \ref{fig5}). The spectra are obtained in a standard manner via the imaginary part of the macroscopic dielectric function \cite{ORR02}.  We use 20 bands and 256 $k$-points for LiF, 10 bands and 256 $k$-points for AlN and Si. All calculations include local field effects. We obtain a very good agreement of the position and strength of the strong bound-exciton peak in LiF compared to experiment, which is also evident from the good agreement between the calculated and fitted screening parameters shown in Fig. \ref{fig2}. For the smaller-gap materials AlN and Si, the excitonic enhancement of the band-edge spectrum is somewhat underestimated (the bound excitons are not shown in Figs. \ref{fig4} and \ref{fig5} since $E_b$ is smaller than the frequency resolution). Here, the excitonic enhancement effects in the BSE and experimental spectra are due to continuum excitons. Compared to RPA, the SXX spectra in Figs. \ref{fig4} and \ref{fig5} give a much better description of the excitonic enhancement effects. 

In conclusion, we propose a very simple nonempirical screening factor for nonlocal exchange, derived as a simplification of BSE. We show that it is easier to derive a good approximation in the many-body framework than developing a better long-ranged xc kernel for TDDFT. Our SXX approach yields exciton binding energies of a wide range of semiconductors and insulators in good agreement with experiment; the performance is consistently better than currently available TDDFT methods. The SXX method works well for the optical spectra of wide-gap materials, and captures continuum excitonic effects in small-gap materials to some extent, although there is still some room for improvement.

The SXX approach constitutes a first step towards a hybrid xc kernel specifically designed for optical properties in periodic
insulators and semiconductors. In this paper we have focused on the long-range behavior of the xc kernel; the next step will be
to match the SXX approach with suitable xc functionals for the short range to capture local-field effects. This should have
minor effects on strongly bound excitons, but is likely to lead to an improvement of the continuum part of the optical
spectrum. Work along these lines is in progress.

We thank Lucia Reining for very helpful discussions. C.U. thanks the ETSF-Palaiseau group for its hospitality and the Ecole Polytechnique for its support during an extended visit in 2014. Z.-h.Y. and C.U. are supported by NSF grant DMR-1408904. F. S. thanks GENCI for computer time (project 0544).

\bibliography{exciton}

\begin{thebibliography}{43}%
\makeatletter
\providecommand \@ifxundefined [1]{%
 \@ifx{#1\undefined}
}%
\providecommand \@ifnum [1]{%
 \ifnum #1\expandafter \@firstoftwo
 \else \expandafter \@secondoftwo
 \fi
}%
\providecommand \@ifx [1]{%
 \ifx #1\expandafter \@firstoftwo
 \else \expandafter \@secondoftwo
 \fi
}%
\providecommand \natexlab [1]{#1}%
\providecommand \enquote  [1]{``#1''}%
\providecommand \bibnamefont  [1]{#1}%
\providecommand \bibfnamefont [1]{#1}%
\providecommand \citenamefont [1]{#1}%
\providecommand \href@noop [0]{\@secondoftwo}%
\providecommand \href [0]{\begingroup \@sanitize@url \@href}%
\providecommand \@href[1]{\@@startlink{#1}\@@href}%
\providecommand \@@href[1]{\endgroup#1\@@endlink}%
\providecommand \@sanitize@url [0]{\catcode `\\12\catcode `\$12\catcode
  `\&12\catcode `\#12\catcode `\^12\catcode `\_12\catcode `\%12\relax}%
\providecommand \@@startlink[1]{}%
\providecommand \@@endlink[0]{}%
\providecommand \url  [0]{\begingroup\@sanitize@url \@url }%
\providecommand \@url [1]{\endgroup\@href {#1}{\urlprefix }}%
\providecommand \urlprefix  [0]{URL }%
\providecommand \Eprint [0]{\href }%
\providecommand \doibase [0]{http://dx.doi.org/}%
\providecommand \selectlanguage [0]{\@gobble}%
\providecommand \bibinfo  [0]{\@secondoftwo}%
\providecommand \bibfield  [0]{\@secondoftwo}%
\providecommand \translation [1]{[#1]}%
\providecommand \BibitemOpen [0]{}%
\providecommand \bibitemStop [0]{}%
\providecommand \bibitemNoStop [0]{.\EOS\space}%
\providecommand \EOS [0]{\spacefactor3000\relax}%
\providecommand \BibitemShut  [1]{\csname bibitem#1\endcsname}%
\let\auto@bib@innerbib\@empty
\bibitem [{\citenamefont {Hanke}\ and\ \citenamefont {Sham}(1980)}]{HS80}%
  \BibitemOpen
  \bibfield  {author} {\bibinfo {author} {\bibfnamefont {W.}~\bibnamefont
  {Hanke}}\ and\ \bibinfo {author} {\bibfnamefont {L.~J.}\ \bibnamefont
  {Sham}},\ }\href@noop {} {\bibfield  {journal} {\bibinfo  {journal} {Phys.
  Rev. B}\ }\textbf {\bibinfo {volume} {21}},\ \bibinfo {pages} {4656}
  (\bibinfo {year} {1980})}\BibitemShut {NoStop}%
\bibitem [{\citenamefont {Onida}\ \emph {et~al.}(2002)\citenamefont {Onida},
  \citenamefont {Reining},\ and\ \citenamefont {Rubio}}]{ORR02}%
  \BibitemOpen
  \bibfield  {author} {\bibinfo {author} {\bibfnamefont {G.}~\bibnamefont
  {Onida}}, \bibinfo {author} {\bibfnamefont {L.}~\bibnamefont {Reining}}, \
  and\ \bibinfo {author} {\bibfnamefont {A.}~\bibnamefont {Rubio}},\
  }\href@noop {} {\bibfield  {journal} {\bibinfo  {journal} {Rev. Mod. Phys.}\
  }\textbf {\bibinfo {volume} {74}},\ \bibinfo {pages} {601} (\bibinfo {year}
  {2002})}\BibitemShut {NoStop}%
\bibitem [{\citenamefont {Runge}\ and\ \citenamefont {Gross}(1984)}]{RG84}%
  \BibitemOpen
  \bibfield  {author} {\bibinfo {author} {\bibfnamefont {E.}~\bibnamefont
  {Runge}}\ and\ \bibinfo {author} {\bibfnamefont {E.~K.~U.}\ \bibnamefont
  {Gross}},\ }\href@noop {} {\bibfield  {journal} {\bibinfo  {journal} {Phys.
  Rev. Lett.}\ }\textbf {\bibinfo {volume} {52}},\ \bibinfo {pages} {997}
  (\bibinfo {year} {1984})}\BibitemShut {NoStop}%
\bibitem [{\citenamefont {Marques}\ \emph {et~al.}(2012)\citenamefont
  {Marques}, \citenamefont {Maitra}, \citenamefont {Nogueira}, \citenamefont
  {Gross},\ and\ \citenamefont {Rubio}}]{MMNG12}%
  \BibitemOpen
  \bibinfo {editor} {\bibfnamefont {M.~A.~L.}\ \bibnamefont {Marques}},
  \bibinfo {editor} {\bibfnamefont {N.~T.}\ \bibnamefont {Maitra}}, \bibinfo
  {editor} {\bibfnamefont {F.~M.~S.}\ \bibnamefont {Nogueira}}, \bibinfo
  {editor} {\bibfnamefont {E.~K.~U.}\ \bibnamefont {Gross}}, \ and\ \bibinfo
  {editor} {\bibfnamefont {A.}~\bibnamefont {Rubio}},\ eds.,\ \href@noop {}
  {\emph {\bibinfo {title} {Fundamentals of time-dependent density functional
  theory}}},\ Lecture notes in physics\ (\bibinfo  {publisher} {Springer},\
  \bibinfo {address} {Berlin},\ \bibinfo {year} {2012})\BibitemShut {NoStop}%
\bibitem [{\citenamefont {Ullrich}(2012)}]{U12}%
  \BibitemOpen
  \bibfield  {author} {\bibinfo {author} {\bibfnamefont {C.~A.}\ \bibnamefont
  {Ullrich}},\ }\href@noop {} {\emph {\bibinfo {title} {Time-dependent
  density-functional theory: concepts and applications}}}\ (\bibinfo
  {publisher} {Oxford University Press},\ \bibinfo {address} {Oxford},\
  \bibinfo {year} {2012})\BibitemShut {NoStop}%
\bibitem [{\citenamefont {Sottile}\ \emph
  {et~al.}(2003{\natexlab{a}})\citenamefont {Sottile}, \citenamefont
  {Karlsson}, \citenamefont {Reining},\ and\ \citenamefont
  {Aryasetiawan}}]{Sottile2003}%
  \BibitemOpen
  \bibfield  {author} {\bibinfo {author} {\bibfnamefont {F.}~\bibnamefont
  {Sottile}}, \bibinfo {author} {\bibfnamefont {K.}~\bibnamefont {Karlsson}},
  \bibinfo {author} {\bibfnamefont {L.}~\bibnamefont {Reining}}, \ and\
  \bibinfo {author} {\bibfnamefont {F.}~\bibnamefont {Aryasetiawan}},\
  }\href@noop {} {\bibfield  {journal} {\bibinfo  {journal} {Phys. Rev. B}\
  }\textbf {\bibinfo {volume} {68}},\ \bibinfo {pages} {205112} (\bibinfo
  {year} {2003}{\natexlab{a}})}\BibitemShut {NoStop}%
\bibitem [{\citenamefont {Botti}\ \emph {et~al.}(2004)\citenamefont {Botti},
  \citenamefont {Sottile}, \citenamefont {Vast}, \citenamefont {Olevano},
  \citenamefont {Reining}, \citenamefont {Weissker}, \citenamefont {Rubio},
  \citenamefont {Onida}, \citenamefont {Del~Sole},\ and\ \citenamefont
  {Godby}}]{BSVO04}%
  \BibitemOpen
  \bibfield  {author} {\bibinfo {author} {\bibfnamefont {S.}~\bibnamefont
  {Botti}}, \bibinfo {author} {\bibfnamefont {F.}~\bibnamefont {Sottile}},
  \bibinfo {author} {\bibfnamefont {N.}~\bibnamefont {Vast}}, \bibinfo {author}
  {\bibfnamefont {V.}~\bibnamefont {Olevano}}, \bibinfo {author} {\bibfnamefont
  {L.}~\bibnamefont {Reining}}, \bibinfo {author} {\bibfnamefont {H.-C.}\
  \bibnamefont {Weissker}}, \bibinfo {author} {\bibfnamefont {A.}~\bibnamefont
  {Rubio}}, \bibinfo {author} {\bibfnamefont {G.}~\bibnamefont {Onida}},
  \bibinfo {author} {\bibfnamefont {R.}~\bibnamefont {Del~Sole}}, \ and\
  \bibinfo {author} {\bibfnamefont {R.~W.}\ \bibnamefont {Godby}},\ }\href@noop
  {} {\bibfield  {journal} {\bibinfo  {journal} {Phys. Rev. B}\ }\textbf
  {\bibinfo {volume} {69}},\ \bibinfo {pages} {155112} (\bibinfo {year}
  {2004})}\BibitemShut {NoStop}%
\bibitem [{\citenamefont {Sharma}\ \emph {et~al.}(2011)\citenamefont {Sharma},
  \citenamefont {Dewhurst}, \citenamefont {Sanna},\ and\ \citenamefont
  {Gross}}]{SDSG11}%
  \BibitemOpen
  \bibfield  {author} {\bibinfo {author} {\bibfnamefont {S.}~\bibnamefont
  {Sharma}}, \bibinfo {author} {\bibfnamefont {J.~K.}\ \bibnamefont
  {Dewhurst}}, \bibinfo {author} {\bibfnamefont {A.}~\bibnamefont {Sanna}}, \
  and\ \bibinfo {author} {\bibfnamefont {E.~K.~U.}\ \bibnamefont {Gross}},\
  }\href@noop {} {\bibfield  {journal} {\bibinfo  {journal} {Phys. Rev. Lett.}\
  }\textbf {\bibinfo {volume} {107}},\ \bibinfo {pages} {186401} (\bibinfo
  {year} {2011})}\BibitemShut {NoStop}%
\bibitem [{\citenamefont {Trevisanutto}\ \emph {et~al.}(2013)\citenamefont
  {Trevisanutto}, \citenamefont {Terentjevs}, \citenamefont {Constantin},
  \citenamefont {Olevano},\ and\ \citenamefont {Della~Sala}}]{TTCO13}%
  \BibitemOpen
  \bibfield  {author} {\bibinfo {author} {\bibfnamefont {P.~E.}\ \bibnamefont
  {Trevisanutto}}, \bibinfo {author} {\bibfnamefont {A.}~\bibnamefont
  {Terentjevs}}, \bibinfo {author} {\bibfnamefont {L.~A.}\ \bibnamefont
  {Constantin}}, \bibinfo {author} {\bibfnamefont {V.}~\bibnamefont {Olevano}},
  \ and\ \bibinfo {author} {\bibfnamefont {F.}~\bibnamefont {Della~Sala}},\
  }\href@noop {} {\bibfield  {journal} {\bibinfo  {journal} {Phys. Rev. B}\
  }\textbf {\bibinfo {volume} {87}},\ \bibinfo {pages} {205143} (\bibinfo
  {year} {2013})}\BibitemShut {NoStop}%
\bibitem [{\citenamefont {de~Boeij}\ \emph {et~al.}(2001)\citenamefont
  {de~Boeij}, \citenamefont {Kootstra}, \citenamefont {Berger}, \citenamefont
  {van Leeuwen},\ and\ \citenamefont {Snijders}}]{deBoeij2001}%
  \BibitemOpen
  \bibfield  {author} {\bibinfo {author} {\bibfnamefont {P.~L.}\ \bibnamefont
  {de~Boeij}}, \bibinfo {author} {\bibfnamefont {F.}~\bibnamefont {Kootstra}},
  \bibinfo {author} {\bibfnamefont {J.~A.}\ \bibnamefont {Berger}}, \bibinfo
  {author} {\bibfnamefont {R.}~\bibnamefont {van Leeuwen}}, \ and\ \bibinfo
  {author} {\bibfnamefont {J.~G.}\ \bibnamefont {Snijders}},\ }\href@noop {}
  {\bibfield  {journal} {\bibinfo  {journal} {J. Chem. Phys.}\ }\textbf
  {\bibinfo {volume} {115}},\ \bibinfo {pages} {1995} (\bibinfo {year}
  {2001})}\BibitemShut {NoStop}%
\bibitem [{\citenamefont {Reining}\ \emph {et~al.}(2002)\citenamefont
  {Reining}, \citenamefont {Olevano}, \citenamefont {Rubio},\ and\
  \citenamefont {Onida}}]{Reining2002}%
  \BibitemOpen
  \bibfield  {author} {\bibinfo {author} {\bibfnamefont {L.}~\bibnamefont
  {Reining}}, \bibinfo {author} {\bibfnamefont {V.}~\bibnamefont {Olevano}},
  \bibinfo {author} {\bibfnamefont {A.}~\bibnamefont {Rubio}}, \ and\ \bibinfo
  {author} {\bibfnamefont {G.}~\bibnamefont {Onida}},\ }\href@noop {}
  {\bibfield  {journal} {\bibinfo  {journal} {Phys. Rev. Lett.}\ }\textbf
  {\bibinfo {volume} {88}},\ \bibinfo {pages} {066404} (\bibinfo {year}
  {2002})}\BibitemShut {NoStop}%
\bibitem [{\citenamefont {Sottile}\ \emph
  {et~al.}(2003{\natexlab{b}})\citenamefont {Sottile}, \citenamefont
  {Olevano},\ and\ \citenamefont {Reining}}]{SOR03}%
  \BibitemOpen
  \bibfield  {author} {\bibinfo {author} {\bibfnamefont {F.}~\bibnamefont
  {Sottile}}, \bibinfo {author} {\bibfnamefont {V.}~\bibnamefont {Olevano}}, \
  and\ \bibinfo {author} {\bibfnamefont {L.}~\bibnamefont {Reining}},\
  }\href@noop {} {\bibfield  {journal} {\bibinfo  {journal} {Phys. Rev. Lett.}\
  }\textbf {\bibinfo {volume} {91}},\ \bibinfo {pages} {056402} (\bibinfo
  {year} {2003}{\natexlab{b}})}\BibitemShut {NoStop}%
\bibitem [{\citenamefont {Adragna}\ \emph {et~al.}(2003)\citenamefont
  {Adragna}, \citenamefont {Del~Sole},\ and\ \citenamefont {Marini}}]{ADM03}%
  \BibitemOpen
  \bibfield  {author} {\bibinfo {author} {\bibfnamefont {G.}~\bibnamefont
  {Adragna}}, \bibinfo {author} {\bibfnamefont {R.}~\bibnamefont {Del~Sole}}, \
  and\ \bibinfo {author} {\bibfnamefont {A.}~\bibnamefont {Marini}},\
  }\href@noop {} {\bibfield  {journal} {\bibinfo  {journal} {Phys. Rev. B}\
  }\textbf {\bibinfo {volume} {68}},\ \bibinfo {pages} {165108} (\bibinfo
  {year} {2003})}\BibitemShut {NoStop}%
\bibitem [{\citenamefont {Marini}\ \emph {et~al.}(2003)\citenamefont {Marini},
  \citenamefont {Del~Sole},\ and\ \citenamefont {Rubio}}]{MDR03}%
  \BibitemOpen
  \bibfield  {author} {\bibinfo {author} {\bibfnamefont {A.}~\bibnamefont
  {Marini}}, \bibinfo {author} {\bibfnamefont {R.}~\bibnamefont {Del~Sole}}, \
  and\ \bibinfo {author} {\bibfnamefont {A.}~\bibnamefont {Rubio}},\
  }\href@noop {} {\bibfield  {journal} {\bibinfo  {journal} {Phys. Rev. Lett.}\
  }\textbf {\bibinfo {volume} {91}},\ \bibinfo {pages} {256402} (\bibinfo
  {year} {2003})}\BibitemShut {NoStop}%
\bibitem [{\citenamefont {Ghosez}\ \emph {et~al.}(1997)\citenamefont {Ghosez},
  \citenamefont {Gonze},\ and\ \citenamefont {Godby}}]{GGG97}%
  \BibitemOpen
  \bibfield  {author} {\bibinfo {author} {\bibfnamefont {P.}~\bibnamefont
  {Ghosez}}, \bibinfo {author} {\bibfnamefont {X.}~\bibnamefont {Gonze}}, \
  and\ \bibinfo {author} {\bibfnamefont {R.~W.}\ \bibnamefont {Godby}},\
  }\href@noop {} {\bibfield  {journal} {\bibinfo  {journal} {Phys. Rev. B}\
  }\textbf {\bibinfo {volume} {56}},\ \bibinfo {pages} {12811} (\bibinfo {year}
  {1997})}\BibitemShut {NoStop}%
\bibitem [{\citenamefont {Botti}\ \emph {et~al.}(2007)\citenamefont {Botti},
  \citenamefont {Schindlmayr}, \citenamefont {Del~Sole},\ and\ \citenamefont
  {Reining}}]{Botti2007}%
  \BibitemOpen
  \bibfield  {author} {\bibinfo {author} {\bibfnamefont {S.}~\bibnamefont
  {Botti}}, \bibinfo {author} {\bibfnamefont {A.}~\bibnamefont {Schindlmayr}},
  \bibinfo {author} {\bibfnamefont {R.}~\bibnamefont {Del~Sole}}, \ and\
  \bibinfo {author} {\bibfnamefont {L.}~\bibnamefont {Reining}},\ }\href@noop
  {} {\bibfield  {journal} {\bibinfo  {journal} {Rep. Prog. Phys.}\ }\textbf
  {\bibinfo {volume} {70}},\ \bibinfo {pages} {357} (\bibinfo {year}
  {2007})}\BibitemShut {NoStop}%
\bibitem [{\citenamefont {Kim}\ and\ \citenamefont
  {G{\"o}rling}(2002{\natexlab{a}})}]{Kim2002a}%
  \BibitemOpen
  \bibfield  {author} {\bibinfo {author} {\bibfnamefont {Y.-H.}\ \bibnamefont
  {Kim}}\ and\ \bibinfo {author} {\bibfnamefont {A.}~\bibnamefont
  {G{\"o}rling}},\ }\href@noop {} {\bibfield  {journal} {\bibinfo  {journal}
  {Phys. Rev. Lett.}\ }\textbf {\bibinfo {volume} {89}},\ \bibinfo {pages}
  {096402} (\bibinfo {year} {2002}{\natexlab{a}})}\BibitemShut {NoStop}%
\bibitem [{\citenamefont {Kim}\ and\ \citenamefont
  {G{\"o}rling}(2002{\natexlab{b}})}]{Kim2002b}%
  \BibitemOpen
  \bibfield  {author} {\bibinfo {author} {\bibfnamefont {Y.-H.}\ \bibnamefont
  {Kim}}\ and\ \bibinfo {author} {\bibfnamefont {A.}~\bibnamefont
  {G{\"o}rling}},\ }\href@noop {} {\bibfield  {journal} {\bibinfo  {journal}
  {Phys. Rev. B}\ }\textbf {\bibinfo {volume} {66}},\ \bibinfo {pages} {035114}
  (\bibinfo {year} {2002}{\natexlab{b}})}\BibitemShut {NoStop}%
\bibitem [{\citenamefont {Bruneval}\ \emph
  {et~al.}(2006{\natexlab{a}})\citenamefont {Bruneval}, \citenamefont
  {Sottile}, \citenamefont {Olevano},\ and\ \citenamefont
  {Reining}}]{Bruneval2006}%
  \BibitemOpen
  \bibfield  {author} {\bibinfo {author} {\bibfnamefont {F.}~\bibnamefont
  {Bruneval}}, \bibinfo {author} {\bibfnamefont {F.}~\bibnamefont {Sottile}},
  \bibinfo {author} {\bibfnamefont {V.}~\bibnamefont {Olevano}}, \ and\
  \bibinfo {author} {\bibfnamefont {L.}~\bibnamefont {Reining}},\ }\href@noop
  {} {\bibfield  {journal} {\bibinfo  {journal} {J. Chem. Phys.}\ }\textbf
  {\bibinfo {volume} {124}},\ \bibinfo {pages} {144113} (\bibinfo {year}
  {2006}{\natexlab{a}})}\BibitemShut {NoStop}%
\bibitem [{\citenamefont {Stephens}\ \emph {et~al.}(1994)\citenamefont
  {Stephens}, \citenamefont {Devlin}, \citenamefont {Chabalowski},\ and\
  \citenamefont {Frisch}}]{SDCF94}%
  \BibitemOpen
  \bibfield  {author} {\bibinfo {author} {\bibfnamefont {P.~J.}\ \bibnamefont
  {Stephens}}, \bibinfo {author} {\bibfnamefont {F.~J.}\ \bibnamefont
  {Devlin}}, \bibinfo {author} {\bibfnamefont {C.~F.}\ \bibnamefont
  {Chabalowski}}, \ and\ \bibinfo {author} {\bibfnamefont {M.~J.}\ \bibnamefont
  {Frisch}},\ }\href@noop {} {\bibfield  {journal} {\bibinfo  {journal} {J.
  Phys. Chem.}\ }\textbf {\bibinfo {volume} {98}},\ \bibinfo {pages} {11623}
  (\bibinfo {year} {1994})}\BibitemShut {NoStop}%
\bibitem [{\citenamefont {Bernasconi}\ \emph {et~al.}(2011)\citenamefont
  {Bernasconi}, \citenamefont {Tomi\'c}, \citenamefont {Ferrero}, \citenamefont
  {R\'erat}, \citenamefont {Orlando}, \citenamefont {Dovesi},\ and\
  \citenamefont {Harrison}}]{Bernasconi2011}%
  \BibitemOpen
  \bibfield  {author} {\bibinfo {author} {\bibfnamefont {L.}~\bibnamefont
  {Bernasconi}}, \bibinfo {author} {\bibfnamefont {S.}~\bibnamefont {Tomi\'c}},
  \bibinfo {author} {\bibfnamefont {M.}~\bibnamefont {Ferrero}}, \bibinfo
  {author} {\bibfnamefont {M.}~\bibnamefont {R\'erat}}, \bibinfo {author}
  {\bibfnamefont {R.}~\bibnamefont {Orlando}}, \bibinfo {author} {\bibfnamefont
  {R.}~\bibnamefont {Dovesi}}, \ and\ \bibinfo {author} {\bibfnamefont {N.~M.}\
  \bibnamefont {Harrison}},\ }\href@noop {} {\bibfield  {journal} {\bibinfo
  {journal} {Phys. Rev. B}\ }\textbf {\bibinfo {volume} {83}},\ \bibinfo
  {pages} {195325} (\bibinfo {year} {2011})}\BibitemShut {NoStop}%
\bibitem [{\citenamefont {Tomi\'c}\ \emph {et~al.}(2014)\citenamefont
  {Tomi\'c}, \citenamefont {Bernasconi}, \citenamefont {Searle},\ and\
  \citenamefont {Harrison}}]{Tomic2014}%
  \BibitemOpen
  \bibfield  {author} {\bibinfo {author} {\bibfnamefont {S.}~\bibnamefont
  {Tomi\'c}}, \bibinfo {author} {\bibfnamefont {L.}~\bibnamefont {Bernasconi}},
  \bibinfo {author} {\bibfnamefont {B.~G.}\ \bibnamefont {Searle}}, \ and\
  \bibinfo {author} {\bibfnamefont {N.~M.}\ \bibnamefont {Harrison}},\
  }\href@noop {} {\bibfield  {journal} {\bibinfo  {journal} {J. Phys. Chem. C}\
  }\textbf {\bibinfo {volume} {118}},\ \bibinfo {pages} {14478} (\bibinfo
  {year} {2014})}\BibitemShut {NoStop}%
\bibitem [{\citenamefont {Heyd}\ \emph {et~al.}(2003)\citenamefont {Heyd},
  \citenamefont {Scuseria},\ and\ \citenamefont {Ernzerhof}}]{Heyd2003}%
  \BibitemOpen
  \bibfield  {author} {\bibinfo {author} {\bibfnamefont {J.}~\bibnamefont
  {Heyd}}, \bibinfo {author} {\bibfnamefont {G.~E.}\ \bibnamefont {Scuseria}},
  \ and\ \bibinfo {author} {\bibfnamefont {M.}~\bibnamefont {Ernzerhof}},\
  }\href@noop {} {\bibfield  {journal} {\bibinfo  {journal} {J. Chem. Phys.}\
  }\textbf {\bibinfo {volume} {118}},\ \bibinfo {pages} {8207} (\bibinfo {year}
  {2003})}\BibitemShut {NoStop}%
\bibitem [{\citenamefont {Heyd}\ \emph {et~al.}(2006)\citenamefont {Heyd},
  \citenamefont {Scuseria},\ and\ \citenamefont {Ernzerhof}}]{Heyd2006}%
  \BibitemOpen
  \bibfield  {author} {\bibinfo {author} {\bibfnamefont {J.}~\bibnamefont
  {Heyd}}, \bibinfo {author} {\bibfnamefont {G.~E.}\ \bibnamefont {Scuseria}},
  \ and\ \bibinfo {author} {\bibfnamefont {M.}~\bibnamefont {Ernzerhof}},\
  }\href@noop {} {\bibfield  {journal} {\bibinfo  {journal} {J. Chem. Phys.}\
  }\textbf {\bibinfo {volume} {124}},\ \bibinfo {pages} {219906} (\bibinfo
  {year} {2006})}\BibitemShut {NoStop}%
\bibitem [{\citenamefont {Heyd}\ \emph {et~al.}(2005)\citenamefont {Heyd},
  \citenamefont {Peralta}, \citenamefont {Scuseria},\ and\ \citenamefont
  {Martin}}]{Heyd2005}%
  \BibitemOpen
  \bibfield  {author} {\bibinfo {author} {\bibfnamefont {J.}~\bibnamefont
  {Heyd}}, \bibinfo {author} {\bibfnamefont {J.~E.}\ \bibnamefont {Peralta}},
  \bibinfo {author} {\bibfnamefont {G.~E.}\ \bibnamefont {Scuseria}}, \ and\
  \bibinfo {author} {\bibfnamefont {R.~L.}\ \bibnamefont {Martin}},\
  }\href@noop {} {\bibfield  {journal} {\bibinfo  {journal} {J. Chem. Phys.}\
  }\textbf {\bibinfo {volume} {123}},\ \bibinfo {pages} {174101} (\bibinfo
  {year} {2005})}\BibitemShut {NoStop}%
\bibitem [{\citenamefont {Schimka}\ \emph {et~al.}(2011)\citenamefont
  {Schimka}, \citenamefont {Harl},\ and\ \citenamefont {Kresse}}]{Schimka2011}%
  \BibitemOpen
  \bibfield  {author} {\bibinfo {author} {\bibfnamefont {L.}~\bibnamefont
  {Schimka}}, \bibinfo {author} {\bibfnamefont {J.}~\bibnamefont {Harl}}, \
  and\ \bibinfo {author} {\bibfnamefont {G.}~\bibnamefont {Kresse}},\
  }\href@noop {} {\bibfield  {journal} {\bibinfo  {journal} {J. Chem. Phys.}\
  }\textbf {\bibinfo {volume} {134}},\ \bibinfo {pages} {024116} (\bibinfo
  {year} {2011})}\BibitemShut {NoStop}%
\bibitem [{\citenamefont {Moussa}\ \emph {et~al.}(2012)\citenamefont {Moussa},
  \citenamefont {Schultz},\ and\ \citenamefont {Chelikowsky}}]{Moussa2012}%
  \BibitemOpen
  \bibfield  {author} {\bibinfo {author} {\bibfnamefont {J.~E.}\ \bibnamefont
  {Moussa}}, \bibinfo {author} {\bibfnamefont {P.~A.}\ \bibnamefont {Schultz}},
  \ and\ \bibinfo {author} {\bibfnamefont {J.~R.}\ \bibnamefont
  {Chelikowsky}},\ }\href@noop {} {\bibfield  {journal} {\bibinfo  {journal}
  {J. Chem. Phys.}\ }\textbf {\bibinfo {volume} {136}},\ \bibinfo {pages}
  {204117} (\bibinfo {year} {2012})}\BibitemShut {NoStop}%
\bibitem [{\citenamefont {Paier}\ \emph {et~al.}(2008)\citenamefont {Paier},
  \citenamefont {Marsman},\ and\ \citenamefont {Kresse}}]{Paier2008}%
  \BibitemOpen
  \bibfield  {author} {\bibinfo {author} {\bibfnamefont {J.}~\bibnamefont
  {Paier}}, \bibinfo {author} {\bibfnamefont {M.}~\bibnamefont {Marsman}}, \
  and\ \bibinfo {author} {\bibfnamefont {G.}~\bibnamefont {Kresse}},\
  }\href@noop {} {\bibfield  {journal} {\bibinfo  {journal} {Phys. Rev. B}\
  }\textbf {\bibinfo {volume} {78}},\ \bibinfo {pages} {121201} (\bibinfo
  {year} {2008})}\BibitemShut {NoStop}%
\bibitem [{\citenamefont {Casida}(1996)}]{C96}%
  \BibitemOpen
  \bibfield  {author} {\bibinfo {author} {\bibfnamefont {M.~E.}\ \bibnamefont
  {Casida}},\ }in\ \href@noop {} {\emph {\bibinfo {booktitle} {Recent
  developments and applications in density functional theory}}},\ \bibinfo
  {editor} {edited by\ \bibinfo {editor} {\bibfnamefont {J.~M.}\ \bibnamefont
  {Seminario}}}\ (\bibinfo  {publisher} {Elsevier, Amsterdam},\ \bibinfo {year}
  {1996})\BibitemShut {NoStop}%
\bibitem [{\citenamefont {Ullrich}\ and\ \citenamefont {Yang}(2014)}]{UY14}%
  \BibitemOpen
  \bibfield  {author} {\bibinfo {author} {\bibfnamefont {C.~A.}\ \bibnamefont
  {Ullrich}}\ and\ \bibinfo {author} {\bibfnamefont {Z.-H.}\ \bibnamefont
  {Yang}},\ }\href@noop {} {\bibfield  {journal} {\bibinfo  {journal}
  {Brazilian J. Phys.}\ }\textbf {\bibinfo {volume} {44}},\ \bibinfo {pages}
  {154} (\bibinfo {year} {2014})}\BibitemShut {NoStop}%
\bibitem [{\citenamefont {Bruneval}\ \emph
  {et~al.}(2006{\natexlab{b}})\citenamefont {Bruneval}, \citenamefont
  {Sottile}, \citenamefont {Olevano},\ and\ \citenamefont {Reining}}]{BSOR06}%
  \BibitemOpen
  \bibfield  {author} {\bibinfo {author} {\bibfnamefont {F.}~\bibnamefont
  {Bruneval}}, \bibinfo {author} {\bibfnamefont {F.}~\bibnamefont {Sottile}},
  \bibinfo {author} {\bibfnamefont {V.}~\bibnamefont {Olevano}}, \ and\
  \bibinfo {author} {\bibfnamefont {L.}~\bibnamefont {Reining}},\ }\href@noop
  {} {\bibfield  {journal} {\bibinfo  {journal} {J. Chem. Phys.}\ }\textbf
  {\bibinfo {volume} {124}},\ \bibinfo {pages} {144113} (\bibinfo {year}
  {2006}{\natexlab{b}})}\BibitemShut {NoStop}%
\bibitem [{\citenamefont {Yang}\ and\ \citenamefont {Ullrich}(2013)}]{YU13}%
  \BibitemOpen
  \bibfield  {author} {\bibinfo {author} {\bibfnamefont {Z.-H.}\ \bibnamefont
  {Yang}}\ and\ \bibinfo {author} {\bibfnamefont {C.~A.}\ \bibnamefont
  {Ullrich}},\ }\href@noop {} {\bibfield  {journal} {\bibinfo  {journal} {Phys.
  Rev. B}\ }\textbf {\bibinfo {volume} {87}},\ \bibinfo {pages} {195204}
  (\bibinfo {year} {2013})}\BibitemShut {NoStop}%
\bibitem [{\citenamefont {Yang}\ \emph {et~al.}(2012)\citenamefont {Yang},
  \citenamefont {Li},\ and\ \citenamefont {Ullrich}}]{YLU12}%
  \BibitemOpen
  \bibfield  {author} {\bibinfo {author} {\bibfnamefont {Z.-H.}\ \bibnamefont
  {Yang}}, \bibinfo {author} {\bibfnamefont {Y.}~\bibnamefont {Li}}, \ and\
  \bibinfo {author} {\bibfnamefont {C.~A.}\ \bibnamefont {Ullrich}},\
  }\href@noop {} {\bibfield  {journal} {\bibinfo  {journal} {J. Chem. Phys.}\
  }\textbf {\bibinfo {volume} {137}},\ \bibinfo {pages} {014513} (\bibinfo
  {year} {2012})}\BibitemShut {NoStop}%
\bibitem [{\citenamefont {Marques}\ \emph {et~al.}(2011)\citenamefont
  {Marques}, \citenamefont {Vidal}, \citenamefont {Oliveira}, \citenamefont
  {Reining},\ and\ \citenamefont {Botti}}]{MVOR11}%
  \BibitemOpen
  \bibfield  {author} {\bibinfo {author} {\bibfnamefont {M.~A.~L.}\
  \bibnamefont {Marques}}, \bibinfo {author} {\bibfnamefont {J.}~\bibnamefont
  {Vidal}}, \bibinfo {author} {\bibfnamefont {M.~J.~T.}\ \bibnamefont
  {Oliveira}}, \bibinfo {author} {\bibfnamefont {L.}~\bibnamefont {Reining}}, \
  and\ \bibinfo {author} {\bibfnamefont {S.}~\bibnamefont {Botti}},\
  }\href@noop {} {\bibfield  {journal} {\bibinfo  {journal} {Phys. Rev. B}\
  }\textbf {\bibinfo {volume} {83}},\ \bibinfo {pages} {035119} (\bibinfo
  {year} {2011})}\BibitemShut {NoStop}%
\bibitem [{\citenamefont {Benedict}\ \emph {et~al.}(1998)\citenamefont
  {Benedict}, \citenamefont {Shirley},\ and\ \citenamefont
  {Bohn}}]{Shirley1998}%
  \BibitemOpen
  \bibfield  {author} {\bibinfo {author} {\bibfnamefont {L.~X.}\ \bibnamefont
  {Benedict}}, \bibinfo {author} {\bibfnamefont {E.~L.}\ \bibnamefont
  {Shirley}}, \ and\ \bibinfo {author} {\bibfnamefont {R.~B.}\ \bibnamefont
  {Bohn}},\ }\href@noop {} {\bibfield  {journal} {\bibinfo  {journal} {Phys.
  Rev. B}\ }\textbf {\bibinfo {volume} {57}},\ \bibinfo {pages} {R9385}
  (\bibinfo {year} {1998})}\BibitemShut {NoStop}%
\bibitem [{\citenamefont {Hedin}(1965)}]{Hedin1965}%
  \BibitemOpen
  \bibfield  {author} {\bibinfo {author} {\bibfnamefont {L.}~\bibnamefont
  {Hedin}},\ }\href@noop {} {\bibfield  {journal} {\bibinfo  {journal} {Phys.
  Rev.}\ }\textbf {\bibinfo {volume} {139}},\ \bibinfo {pages} {796} (\bibinfo
  {year} {1965})}\BibitemShut {NoStop}%
\bibitem [{\citenamefont {Cudazzo}\ \emph {et~al.}(2011)\citenamefont
  {Cudazzo}, \citenamefont {Tokatly},\ and\ \citenamefont {Rubio}}]{cudazzo}%
  \BibitemOpen
  \bibfield  {author} {\bibinfo {author} {\bibfnamefont {P.}~\bibnamefont
  {Cudazzo}}, \bibinfo {author} {\bibfnamefont {I.~V.}\ \bibnamefont
  {Tokatly}}, \ and\ \bibinfo {author} {\bibfnamefont {A.}~\bibnamefont
  {Rubio}},\ }\href {http://link.aps.org/doi/10.1103/PhysRevB.84.085406}
  {\bibfield  {journal} {\bibinfo  {journal} {Phys. Rev. B}\ }\textbf {\bibinfo
  {volume} {84}},\ \bibinfo {pages} {085406} (\bibinfo {year}
  {2011})}\BibitemShut {NoStop}%
\bibitem [{\citenamefont {Gonze}\ \emph {et~al.}(2009)\citenamefont {Gonze},
  \citenamefont {Amadon}, \citenamefont {Anglade}, \citenamefont {Beuken},
  \citenamefont {Bottin}, \citenamefont {Boulanger}, \citenamefont {Bruneval},
  \citenamefont {Caliste}, \citenamefont {Caracas}, \citenamefont
  {C{\^ot}{\'e}}, \citenamefont {Deutsch}, \citenamefont {Genovese},
  \citenamefont {Ghosez}, \citenamefont {Giantomassi}, \citenamefont
  {Goedecker}, \citenamefont {Hamann}, \citenamefont {Hermet}, \citenamefont
  {Jollet}, \citenamefont {Jomard}, \citenamefont {Leroux}, \citenamefont
  {Mancini}, \citenamefont {Mazevet}, \citenamefont {Oliveira}, \citenamefont
  {Onida}, \citenamefont {Pouillon}, \citenamefont {Rangel}, \citenamefont
  {Rignanese}, \citenamefont {Sangalli}, \citenamefont {Shaltaf}, \citenamefont
  {Torrent}, \citenamefont {Verstraete}, \citenamefont {Zerah},\ and\
  \citenamefont {Zwanziger}}]{ABINIT}%
  \BibitemOpen
  \bibfield  {author} {\bibinfo {author} {\bibfnamefont {X.}~\bibnamefont
  {Gonze}}, \bibinfo {author} {\bibfnamefont {B.}~\bibnamefont {Amadon}},
  \bibinfo {author} {\bibfnamefont {P.-M.}\ \bibnamefont {Anglade}}, \bibinfo
  {author} {\bibfnamefont {J.-M.}\ \bibnamefont {Beuken}}, \bibinfo {author}
  {\bibfnamefont {F.}~\bibnamefont {Bottin}}, \bibinfo {author} {\bibfnamefont
  {P.}~\bibnamefont {Boulanger}}, \bibinfo {author} {\bibfnamefont
  {F.}~\bibnamefont {Bruneval}}, \bibinfo {author} {\bibfnamefont
  {D.}~\bibnamefont {Caliste}}, \bibinfo {author} {\bibfnamefont
  {R.}~\bibnamefont {Caracas}}, \bibinfo {author} {\bibfnamefont
  {M.}~\bibnamefont {C{\^ot}{\'e}}}, \bibinfo {author} {\bibfnamefont
  {T.}~\bibnamefont {Deutsch}}, \bibinfo {author} {\bibfnamefont
  {L.}~\bibnamefont {Genovese}}, \bibinfo {author} {\bibfnamefont
  {P.}~\bibnamefont {Ghosez}}, \bibinfo {author} {\bibfnamefont
  {M.}~\bibnamefont {Giantomassi}}, \bibinfo {author} {\bibfnamefont
  {S.}~\bibnamefont {Goedecker}}, \bibinfo {author} {\bibfnamefont {D.~R.}\
  \bibnamefont {Hamann}}, \bibinfo {author} {\bibfnamefont {P.}~\bibnamefont
  {Hermet}}, \bibinfo {author} {\bibfnamefont {F.}~\bibnamefont {Jollet}},
  \bibinfo {author} {\bibfnamefont {G.}~\bibnamefont {Jomard}}, \bibinfo
  {author} {\bibfnamefont {S.}~\bibnamefont {Leroux}}, \bibinfo {author}
  {\bibfnamefont {M.}~\bibnamefont {Mancini}}, \bibinfo {author} {\bibfnamefont
  {S.}~\bibnamefont {Mazevet}}, \bibinfo {author} {\bibfnamefont {M.~J.~T.}\
  \bibnamefont {Oliveira}}, \bibinfo {author} {\bibfnamefont {G.}~\bibnamefont
  {Onida}}, \bibinfo {author} {\bibfnamefont {Y.}~\bibnamefont {Pouillon}},
  \bibinfo {author} {\bibfnamefont {T.}~\bibnamefont {Rangel}}, \bibinfo
  {author} {\bibfnamefont {G.-M.}\ \bibnamefont {Rignanese}}, \bibinfo {author}
  {\bibfnamefont {D.}~\bibnamefont {Sangalli}}, \bibinfo {author}
  {\bibfnamefont {R.}~\bibnamefont {Shaltaf}}, \bibinfo {author} {\bibfnamefont
  {M.}~\bibnamefont {Torrent}}, \bibinfo {author} {\bibfnamefont {M.~J.}\
  \bibnamefont {Verstraete}}, \bibinfo {author} {\bibfnamefont
  {G.}~\bibnamefont {Zerah}}, \ and\ \bibinfo {author} {\bibfnamefont {J.~W.}\
  \bibnamefont {Zwanziger}},\ }\href@noop {} {\bibfield  {journal} {\bibinfo
  {journal} {Computer Phys. Comm.}\ }\textbf {\bibinfo {volume} {180}},\
  \bibinfo {pages} {2582} (\bibinfo {year} {2009})}\BibitemShut {NoStop}%
\bibitem [{\citenamefont {Monkhorst}\ and\ \citenamefont
  {Pack}(1976)}]{Monkhorst1976}%
  \BibitemOpen
  \bibfield  {author} {\bibinfo {author} {\bibfnamefont {H.~J.}\ \bibnamefont
  {Monkhorst}}\ and\ \bibinfo {author} {\bibfnamefont {J.~D.}\ \bibnamefont
  {Pack}},\ }\href@noop {} {\bibfield  {journal} {\bibinfo  {journal} {Phys.
  Rev. B}\ }\textbf {\bibinfo {volume} {13}},\ \bibinfo {pages} {5188}
  (\bibinfo {year} {1976})}\BibitemShut {NoStop}%
\bibitem [{\citenamefont {Roessler}\ and\ \citenamefont {Walker}(1967)}]{RW67}%
  \BibitemOpen
  \bibfield  {author} {\bibinfo {author} {\bibfnamefont {D.~M.}\ \bibnamefont
  {Roessler}}\ and\ \bibinfo {author} {\bibfnamefont {W.~C.}\ \bibnamefont
  {Walker}},\ }\href@noop {} {\bibfield  {journal} {\bibinfo  {journal} {J.
  Opt. Soc. Am.}\ }\textbf {\bibinfo {volume} {57}},\ \bibinfo {pages} {835}
  (\bibinfo {year} {1967})}\BibitemShut {NoStop}%
\bibitem [{\citenamefont {Cimalla}\ \emph {et~al.}(2005)\citenamefont
  {Cimalla}, \citenamefont {Lebedev}, \citenamefont {Kaiser}, \citenamefont
  {Goldhahn}, \citenamefont {Foerster}, \citenamefont {Pezoldt},\ and\
  \citenamefont {Ambacher}}]{CLKG05}%
  \BibitemOpen
  \bibfield  {author} {\bibinfo {author} {\bibfnamefont {V.}~\bibnamefont
  {Cimalla}}, \bibinfo {author} {\bibfnamefont {V.}~\bibnamefont {Lebedev}},
  \bibinfo {author} {\bibfnamefont {U.}~\bibnamefont {Kaiser}}, \bibinfo
  {author} {\bibfnamefont {R.}~\bibnamefont {Goldhahn}}, \bibinfo {author}
  {\bibfnamefont {C.}~\bibnamefont {Foerster}}, \bibinfo {author}
  {\bibfnamefont {J.}~\bibnamefont {Pezoldt}}, \ and\ \bibinfo {author}
  {\bibfnamefont {O.}~\bibnamefont {Ambacher}},\ }\href@noop {} {\bibfield
  {journal} {\bibinfo  {journal} {Phys. Status Solidi C}\ }\textbf {\bibinfo
  {volume} {2}},\ \bibinfo {pages} {2199} (\bibinfo {year} {2005})}\BibitemShut
  {NoStop}%
\bibitem [{\citenamefont {Benedict}\ \emph {et~al.}(1999)\citenamefont
  {Benedict}, \citenamefont {Wethkamp}, \citenamefont {Wilmers}, \citenamefont
  {Cobet}, \citenamefont {Esser}, \citenamefont {Shirley}, \citenamefont
  {Richter},\ and\ \citenamefont {Cardona}}]{BWWC99}%
  \BibitemOpen
  \bibfield  {author} {\bibinfo {author} {\bibfnamefont {L.~X.}\ \bibnamefont
  {Benedict}}, \bibinfo {author} {\bibfnamefont {T.}~\bibnamefont {Wethkamp}},
  \bibinfo {author} {\bibfnamefont {K.}~\bibnamefont {Wilmers}}, \bibinfo
  {author} {\bibfnamefont {C.}~\bibnamefont {Cobet}}, \bibinfo {author}
  {\bibfnamefont {N.}~\bibnamefont {Esser}}, \bibinfo {author} {\bibfnamefont
  {E.~L.}\ \bibnamefont {Shirley}}, \bibinfo {author} {\bibfnamefont
  {W.}~\bibnamefont {Richter}}, \ and\ \bibinfo {author} {\bibfnamefont
  {M.}~\bibnamefont {Cardona}},\ }\href@noop {} {\bibfield  {journal} {\bibinfo
   {journal} {Solid State Comm.}\ }\textbf {\bibinfo {volume} {112}},\ \bibinfo
  {pages} {129} (\bibinfo {year} {1999})}\BibitemShut {NoStop}%
\bibitem [{\citenamefont {Lautenschlager}\ \emph {et~al.}(1987)\citenamefont
  {Lautenschlager}, \citenamefont {Garriga}, \citenamefont {Vi\~{n}a},\ and\
  \citenamefont {Cardona}}]{LGVC87}%
  \BibitemOpen
  \bibfield  {author} {\bibinfo {author} {\bibfnamefont {P.}~\bibnamefont
  {Lautenschlager}}, \bibinfo {author} {\bibfnamefont {M.}~\bibnamefont
  {Garriga}}, \bibinfo {author} {\bibfnamefont {L.}~\bibnamefont {Vi\~{n}a}}, \
  and\ \bibinfo {author} {\bibfnamefont {M.}~\bibnamefont {Cardona}},\
  }\href@noop {} {\bibfield  {journal} {\bibinfo  {journal} {Phys. Rev. B}\
  }\textbf {\bibinfo {volume} {36}},\ \bibinfo {pages} {4821} (\bibinfo {year}
  {1987})}\BibitemShut {NoStop}%
\end{thebibliography}%

\end{document}